\documentclass[twocolumn,showpacs,preprintnumbers,amsmath,amssymb]{revtex4}
\usepackage{graphicx}
\usepackage{dcolumn}
\usepackage{bm}

\begin{document}

\title{Symplectic structure and monopole strength in $^{12}$C}
\author{T. Yoshida}

\affiliation{
Center for Nuclear Study, Graduate School of Science, University of Tokyo,
7-3-1 Hongo, 133-0033 Tokyo, Japan} 

\author{N. Itagaki}

\affiliation{
Yukawa Institute for Theoretical Physics, Kyoto University,
Kitashirakawa Oiwake-Cho, 606-8502 Kyoto, Japan} 

\author{K. Kat\=o}

\affiliation{
Department of Physics, Faculty of Science, Hokkaido University,
Kita 10 Nishi 8, 060-0810 Sapporo, Japan}
 
\date{\today}

\begin{abstract}
The relation between the monopole transition strength and 
existence of cluster structure in the excited states is discussed
 based on an algebraic cluster model. 
The structure of $^{12}$C is studied with a 3$\alpha$ model, 
and the wave function for the relative motions between $\alpha$ clusters
are described by the symplectic algebra $Sp(2,R)_z$,
which corresponds to the linear combinations
of $SU(3)$ states with different multiplicities.
Introducing $Sp(2,R)_z$ algebra works well for reducing the 
number of the basis states,
and it is also shown that states connected by
the strong monopole transition are classified
by a quantum number $\Lambda$ of the $Sp(2,R)_z$ algebra.
\end{abstract}

\pacs{21.10.-k,21.60.-n,21.60.Gx,27.20.+n,27.30.+t}

\maketitle

\section{Introduction}
Light nuclear systems show many different properties in the structure.
Around the low-lying energy region,
the mean field and the associated shell structure are dominant properties, 
however cluster structures appear close to their decay thresholds.
In this context, an $\alpha$-particle, 
which is strongly bound and an $\alpha$-$\alpha$ interaction is not strong enough 
to make a bound state, can be considered as an effective building block of the structure
of light nuclei \cite{Ikeda}. 
One of the typical examples of cluster structures
is the second $0^+$ ($0^+_2$) state of $^{12}$C at $E_x = 7.65$ MeV
just above the 3$\alpha$-threshold energy.
This state is considered to have
an exotic cluster structure of 3$\alpha$ in analogy with the so-called
``mysterious $0^+$ state" of $^{16}$O at 
$E_x = 6.06$ MeV, which has a $^{12}$C+$\alpha$
cluster structure and is hardly explained 
by a simple shell-model picture.
The $0^+_2$ state plays a crucial role
in synthesis of $^{12}$C from three $^4$He nuclei
in stars \cite{Hoyle}, and the state
has been proven to contain a developed 3$\alpha$-configuration
by many microscopic cluster calculations \cite{Fujiwara, N-dis},
which is a gas-like
state without a specific geometrical-shape.
This state is recently reinterpreted 
as an $\alpha$-condensed state \cite{Cond, Funaki-1, Chernykh}.

To prove the existence of cluster states,
recently it has been proposed that the strong enhancement 
of isoscalar monopole (E0) transitions can be a measure
of the cluster structure \cite{Kawabata}.  
For instance, the presence of the cluster states in $^{13}$C
has been suggested by measuring the isoscalar E0 transitions
from the ground $1/2^-$ state induced by the $^{13}$C$(\alpha, \alpha')^{13}$C
reaction \cite{Sasamoto}. The obtained crosssections are much larger than those
of the shell-model calculations, which suggest that protons
and neutrons are coherently excited and they have spatially
extended distribution in the excited states.

From the theoretical side, 
the relation between the monopole transition strength and the cluster
structure has also been discussed \cite{YIO, Uegaki, HO, Yamada-1}. 
The basic idea arises from the Bayman-Bohr theorem \cite{BB59},
which shows that the lowest representation of the shell-model
contains a component of the lowest $SU(3)$ representation of the cluster states. 
Thus, even cluster states with 
spatially extended distribution,
such as the second $0^+$ state of $^{12}$C, can be generated
by multiplying operators to the shell-model-like ground state.
The monopole operator is the very one which induces the spatial extension
of the ground state and connects it to cluster states
by raising the quanta of the cluster-cluster relative wave function by two.
The monopole matrix element of  $^{12}$C ($0^+_1 \to 0^+_2$)
calculated with the cluster model agrees with the experimental value
(5.4$\pm$0.2 fm$^2$ for proton part \cite{AS-B}), and this is
much larger than that given in the $p$-shell single particle models.
This is one of supports for the proposal that strong monopole transition
can be a signature of 4$N$ correlated states from the theoretical side.
It is also discussed that the mixing of the cluster component in the ground state
is another important factor for the enhancement of the monopole transition
 strength to cluster states \cite{Yamada-1}.

In the present study,
the relation between the monopole transition strength and 
existence of well-developed cluster structure in the excited states is discussed based on
an algebraic cluster model. 
The structure of $^{12}$C is studied with a 3$\alpha$ model, 
and the wave functions for the relative motions between $\alpha$ clusters
are described by 
the harmonic oscillator (HO) basis states forming symplectic algebra.  
The importance of the symplectic structure for light nuclei has been investigated also in \cite{SP2,Arickx}, 
and the relation between the symplectic algebra and the cluster model has been discussed.
 In our study, we focus on the relation between the symplectic structure and monopole transition strength.
As a final goal of this study, we aim to treat the solution of the unbound states 
in a correct way and explicitly impose the boundary conditions 
in outer region. 
For this purpose, it is necessary to introduce basis states with
large principal quantum numbers for the relative motion of clusters,
but the number of the basis states drastically increases
with increasing principal quantum numbers if we adopt $SU(3)$ algebra.

This problem is overcome by introducing symplectic algebra $Sp(2,R)_z$,
where the basis states correspond to the linear combinations
of $SU(3)$ states with different multiplicities.
This $Sp(2,R)_z$ algebra can be a powerful tool to 
create the states corresponding to the excitation modes
of relative motions between $\alpha$ clusters.
The cluster states with 
$SU(3)$ representations which have different total HO
quanta are connected by a common eigenvalue $\Lambda$ of the $Sp(2,R)_z$ algebra,
and it will be shown that
strong monopole transitions are classified by this $\Lambda$.
It is also discussed that limited
values (small values) of $\Lambda$ are enough to achieve good convergence for the
states corresponding to the excitation modes of the clusters \cite{Kato-sp2}.  
Because of this effect, we can adopt states with large values of the HO quanta into 
the model space in this study.

The outline of this paper is given as follows.
Firstly, we show the framework of the symplectic model in sec. II.
In sec. III, we calculate the energy and the monopole transition strength of $^{12}$C.
Here, we discuss the relation between the symplectic quanta $\Lambda$ and the monopole transition strength.
We summarize the discussion in sec. IV.

\section{$Sp(2, R)_z$ basis representation of the 3$\alpha$ model}
We show how to construct a model space 
of the 3$\alpha$ system based on the $SU(3)$ algebra.
However, the $Sp(2,R)_z$ algebra, which corresponds to the linear combination
of $SU(3)$ basis states with different multiplicities, is 
shown to give better description for the cluster states.
The relation between the $SU(3)$ and $Sp(2,R)_z$ 
model spaces is discussed.

\subsection{$SU(3)$ model space}
Here, we show how to construct basis states of the 3$\alpha$-cluster system
 based on the $SU(3)$ algebra.
The $SU(3)$ state of the three-$\alpha$ cluster model for $^{12}$C
 is given by a product of $SU(3)$ states
corresponding to the two Jacobi coordinates 
for the relative motions of $\alpha$-$\alpha$ ($\vec{r}$) and 
($\alpha$-$\alpha$)-$\alpha$ ($\vec{R}$): 
\begin{equation}
SU(3) = SU(3) \otimes SU(3).
\end{equation}
Using the $(\lambda, \mu)\rho$ representation of $SU(3)$,
the basis state with the principal HO quantum numbers $N$ is expressed as
\begin{equation}
N(\lambda, \mu)\rho \sim (N_1, 0) \otimes (N_2, 0),
\end{equation}
where $N_1$ and $N_2$ are principal HO quantum numbers ($N=N_1+N_2$)
for the Jacobi coordinates $(\vec{r}$ and $\vec{R})$
and $\rho$ is the multiplicity of the $(\lambda, \mu)$ state.
Following Refs. \cite{Ka86, Ka88},
the basis function with the values of 
$N(\lambda, \mu)\rho$, $N_1$, $N_2$, $J$ and $K$ is given as,
\begin{eqnarray}
V^{N(\lambda,\mu),J,K}_{N_1,N_2}(\vec{r},\vec{R})&
=&\sum_{l_1,l_2} \langle (N_1, 0), l_1, (N_2, 0), l_2 || N(\lambda,\mu),
J, K \rangle \nonumber \\
& & \times [u_{N_1,l_1}(\vec{r})u_{N_2,l_2}(\vec{R})]_J,
\end{eqnarray}
where $l_1$ and $l_2$ are angular momenta of each Jacobi coordinate,
 $J$ is the total angular momentum and $K$ is the orthonormalized $K$-quantum number of $J$.
We take summation over $N_1$ and $N_2$ in the following way:
\begin{eqnarray}
U^{J^{\pi}}_i(\vec{r}, \vec{R})  & = & \sum_{N_1+N_2=N}A^{N(\lambda,\mu)\rho}_{N_1,N_2}V^{N(\lambda,\mu),J,K}_{N_1,N_2}(r,R), 
\end{eqnarray}
where the index $i$ denotes an abbreviation of $N(\lambda,\mu)\rho,K$. 
In order to take into account the Pauli principle between nuleons belonging to different $\alpha$-clusters,
the coefficients $ A^{N(\lambda,\mu)\rho}_{N_1,N_2}$ 
must be determined by the orthogonal condition model (OCM) \cite{Saito1, Saito2}.
First of all, the value of $N_1$ should be Pauli allowed one ($N_1= 4, 6, 8, \cdots, N$). 
For $N_2$, instead of directly calculating the Pauli allowed state 
for the Jacobi coordinate $\vec{R}$ \cite{KB}, here we calculate the overlap with the Pauli forbidden state of 
rearranged Jacobi coordinates.
Eventually, the Pauli allowed basis states for Jacobi coordinates $(\vec{r}, \vec{R})$ 
are obtained by orthogonalizing the basis states to the 
Pauli forbidden ones with other (rearranged) sets of Jacobi coordinates $(\vec{r'}, \vec{R'})$ and $(\vec{r''},\vec{R''})$.
Here, it is enough if we only consider the Pauli forbidden states for the coordinates $\vec{r'}$ and $\vec{r''}$,
 which have the principal quantum number of $N'_1, N''_1= 0, 1, 2, 3, 5, 7, 9, \cdots$.
This is equivalent to the following condition \cite{Ho5358}, 
\begin{eqnarray}
\hat{Q}|N(\lambda,\mu)k\rangle=q_k|N(\lambda,\mu)k\rangle.
\label{projection}
\end{eqnarray}
Here, the operator $\hat{Q}$ expresses the projection to 
the Pauli forbidden states for all different Jacobi coordinates, and
 the Pauli allowed states are obtained as the eigenstates of $q_k=0$,
 because they have to be orthogonal to all the Pauli forbidden states.
The index $k$ is needed to distinguish the multiplicity of the wave function,
 which has a set of the HO quanta of $N_1$ and $N_2$.
The wave function of the 3$\alpha$ model for $^{12}$C is constructed
 by superposing $U_i(\vec{r},\vec{R})$ basis states.
The size of the model space is determined by the maximum HO quanta $N_{max}$ as follows;
\begin{eqnarray}
\Phi^{J^{\pi}}=\sum_{i=N(\lambda,\mu)\rho,K}c^{J^{\pi}}_{i}U^{J^{\pi}}_i(\vec{r}, \vec{R}),
\end{eqnarray}
where the summation runs under the condition $N\le N_{max}$.
\subsection{Hamiltonian}
The Hamiltonian $H$ is given in the following form:
\begin{eqnarray}
H=T_{\vec{r}}+T_{\vec{R}}+\sum_{i>j}V_{\alpha\alpha}(\vec{r_{ij}})+V^J_{3by}(\vec{r_1},\vec{r_2},\vec{r_3}),
\end{eqnarray}
where $T_{\vec{r}}$ and $T_{\vec{R}}$ are relative kinetic energies corresponding to
 the Jacobi coordinates. 
As for the two-body nuclear interaction, we use the following $\alpha$-$\alpha$ folding potential 
\begin{eqnarray}
V_{\alpha\alpha}(r)=V_2 \exp(-\alpha r^2),
\end{eqnarray}
employed by Kurokawa $et\ al.$ so as to reproduce the observed $\alpha-\alpha$ phase shifts \cite{SW, Kurokawa}.
Here, $\alpha=$0.2009 fm$^{-2}$ and $V_2=-$106.1 MeV are used.
The Coulomb interaction has the following form,
\begin{eqnarray}
V^{\alpha\alpha}_c (r)=\frac{4e^2}{r}\text{erf}(\beta r),
\end{eqnarray}
where $\beta=0.5972$ fm$^{-1}$.
 Moreover, we add an inter three-$\alpha$ interaction:
\begin{eqnarray}
V^J_{3by}(\vec{r_1},\vec{r_2},\vec{r_3}) &=& V^J_3 \exp(-\eta\{r_{12}^2+r_{23}^2+r_{31}^2\}),\label{eqn:3body}
\end{eqnarray}
where $\eta=$0.15 fm$^{-2}$ and $r_{ij}=r_i-r_j$.
In order to reproduce the experimental binding and excitation energies of
 the ground band states ($0^+$, $2^+$ and $4^+$) of $^{12}$C \cite{Kurokawa},
we need to use the strength of the three-body interaction $V^J_3$ as,
 31.7 MeV for $J=0^+$,  63.0 MeV for $J=2^+$ and 150.0 MeV for $J=4^+$, respectively.

Energies and their eigenstates (Eq. (6)) are obtained by diagonalization of the Hamiltonian (Eq. (7)).
In Fig. \ref{fig:sp2rconv}, we show the convergence of the $0^+$ states as a function of the number of basis states (black dotted lines), where $N_{max}$ is gradually increased from 8 to 46 in the $SU(3)$ bases.
It is shown that the ground states has rapid convergence, which indicates the importance of the
 shell-model like configuration. On the other hand, many excited states show slow convergence, which means that
the $SU(3)$ model space is not suitable for the description of the well-developed cluster states in the excited states. 
This is due to the increase of multiplicity 
useless for the convergence as the total HO quanta $N$ increases.

\subsection{$Sp(2, R)_z$ model space}
\begin{figure}[htbp]
\includegraphics[width=8.6cm]{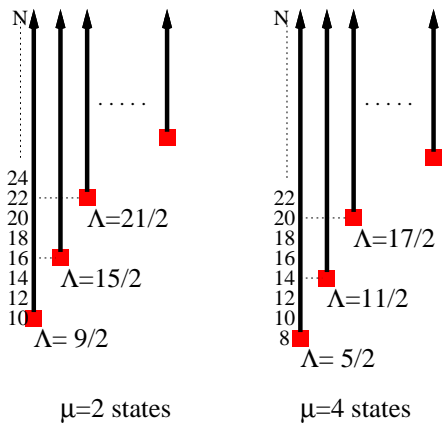}
\caption{
(Color online) Pauli allowed states generated by $Sp(2, R)z$ algebra.
The red squares, vertical lines and arrows show the band head state,
 the principal quantum number $N$ of HO and the ladder states, respectively.
}
\label{fig:sp2rz}
\end{figure}

To achieve the energy convergence in a more efficient way especially for the cluster states in the excited states,
we need appropriate truncation for the model space.
In order to describe the cluster-like configuration, we take into account the major-shell excitation including
 many HO $N$-quanta states.
Here, we intend to correlate different $N$-quanta states by algebraic classifications.
We perform unitary transformation of the states specified by $\rho$ to the other basis sets
by utilizing the $N_1$ and $N_2$ degrees of freedom.
Here, we use the symplectic algebra, $Sp(2, R)_z$.
According to this algebra, basis states are classified by a quantum number $\Lambda$,
which is an eigenvalue of the Casimir operator of this algebra \cite{Kato-sp2}.
This $\Lambda$ specifies the ladder states;
a set of ladder states has definite eigen value of $\Lambda$.
The generators of this algebra are given as
\begin{eqnarray}
\Lambda_+&=&\;\;\;\frac{1}{\sqrt{8}}\sum_{p=1}^{2}a_{z}^{\dag}(p)a_{z}^{\dag}(p), \nonumber\\
\Lambda_-&=&-\frac{1}{\sqrt{8}}\sum_{p=1}^{2}a_{z}(p)a_{z}(p), \nonumber\\
\Lambda_0\;&=&\;\;\;\frac{1}{4}\;\;\sum_{p=1}^{2}(a_{z}^{\dag}(p)a_{z}(p)+a_{z}(p)a_{z}^{\dag}(p)).
\end{eqnarray}
Here, $a_{z}^{\dag}(p)$ and $a_{z}(p)$ are creation and annihilation operator of HO, respectively, where $p$ is an index to distinguish the Jacobi coordinates $\vec{r}$ and $\vec{R}$.
By using these operators, the ladder states are created by multiplying a raising operators $\Lambda_+$ to the band head state, which vanishes when a lowering operator $\Lambda_-$ is multiplied.
Note that each ladder state has a definite eigen value of $\Lambda$, and multiplying $\Lambda_+$ and $\Lambda_-$ does not change this value.
As shown in Fig. \ref{fig:sp2rz}, a new band head state appears when the principal quantum number of HO ($N$) increase by six for each $\mu$ state ($N=\lambda+2\mu$, $\Lambda=\frac{1}{2}(\lambda+\mu)+\frac{1}{4}(n-1)$, where $n$ is an integer).
 However, we need to orthonormalize them by the Gram-Schmidt's procedure, because this new band states are not always orthogonal to the band states which have smaller $\Lambda$ values.
%
\begin{figure}[htbp]
\includegraphics[width=8.6cm]{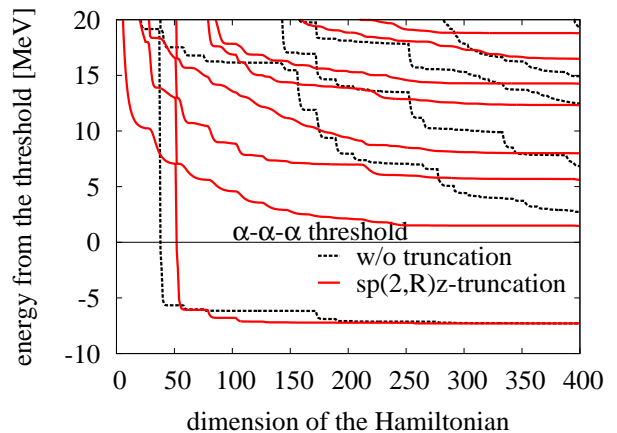}
\caption{(Color online) Energy convergence of the 3$\alpha$ system as a function of the number of basis states.
The black and red lines show the results for the $SU(3)$ and $Sp(2,R)_z$ basis sets, respectively. }
\label{fig:sp2rconv}
\end{figure}

In order to select the model space suited for the description of the excited states, 
we use of the limited $\Lambda$ values.
The truncated model space is expanded by the following bases states as,
\begin{eqnarray}
w^{J^{\pi}}_{\alpha}(\vec{r}, \vec{R})  & = & \sum_{\rho}C^{\Lambda}_{\rho}U^{J^{\pi}}_i(\vec{r}, \vec{R}),
\end{eqnarray}
where the index $\alpha$ denotes an abbreviation of $\Lambda$ and $N(\lambda,\mu)$. 
The equation to be solved is expressed as
\begin{eqnarray}
\sum_{\beta}H_{\alpha,\beta}d^{J^{\pi}}_{k,\beta}=Ed^{J^{\pi}}_{k,\alpha},
\end{eqnarray}
where the matrix element of the Hamiltonian is expressed as,
\begin{eqnarray}
H_{\alpha,\beta}=\langle w_{\alpha}|H|w_{\beta} \rangle.
\end{eqnarray}
The total wave function of the $k$-th state is expressed as
\begin{eqnarray}
\Phi^{J^{\pi}}(k)=\sum_{\alpha}d^{J^{\pi}}_{k,\alpha}w^{J^{\pi}}_{\alpha}(\vec{r}, \vec{R}).
\end{eqnarray}

\begin{figure}[htbp]
\includegraphics[width=8.6cm]{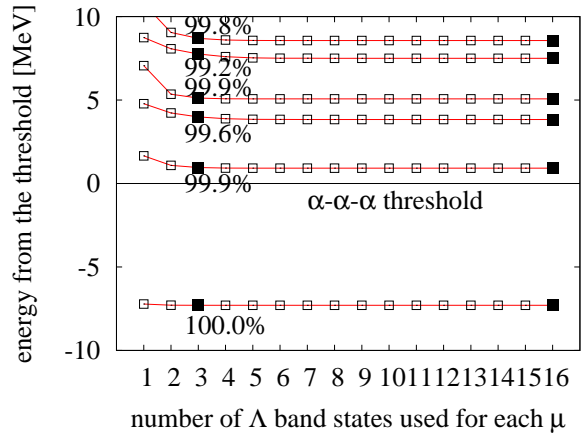}
\caption{(Color online) Energy convergence of the $0^+$ states of the 3$\alpha$ system as a function of the 
 size of the model space (the number of $\Lambda$ band states included in the model space for each $\mu$ state).
The left filled dots show the model space used in the present calculation,
 while the right filled dots show the model space with all $\Lambda$ configurations. The numbers are overlaps with these two wave functions.}
\label{fig:LAMconv}
\end{figure}

We employ this $w^{J^{\pi}}_{\alpha}$ basis set, which is of the $Sp(2,R)_z$ truncation, and 
 shown as the red solid lines of Fig. \ref{fig:sp2rconv}, the energy convergence becomes much faster compared with
the case without this truncation (black dotted lines), especially for the excited states with well-developed cluster configurations.
This good energy convergence can be obtained even if we limit $\Lambda$ values.
Here, we use three lowest $\Lambda$ values for each $\mu$ state.
In return, we take total HO quanta $N_{max}=$ 100 and the $\mu$ values up to 30, 
which is difficult to achieve in $SU(3)$ case.
This gives a model space large enough to describe the cluster states.
In order to confirm the validity of the selection of $\Lambda$ values, we show the energy convergence
 of the $0^+$ states of the 3$\alpha$ system as a function of the size of the model space
 (the number of $\Lambda$ band states included
 in the model space for each $\mu$ state) in Fig. \ref{fig:LAMconv}.
We find that the model space within the three lowest $\Lambda$ values for each $\mu$ state already has enough good convergence (filled points) at this energy region. Moreover, the overlaps between these states and the full $\Lambda$ bands calculation (right filled points) are almost 100\%. 
Therefore, we use this truncated model space in the present calculation.
%

\section{Results}
\subsection{Relation between monopole transitions and $Sp(2,R)_z$ algebra.}
Hereafter we employ a model space in the $Sp(2,R)_z$ representation and
 discuss the relation between the symplectic ladder states and the monopole strengths.
Because the ladder states are created by the operator $\Lambda_+$ ($=\frac{1}{\sqrt{8}}\sum_{p=1}^{2}a_{z}^{\dag}(p)a_{z}^{\dag}(p)$),
it is considered that they have strong relation with the monopole transition,
which is excited by the operator $\hat{E0} \propto r^2+\frac{4}{3}R^2$ with the similar form.

Firstly, we show the ground state properties obtained within the present model space.
The calculated ground state contains the component of the lowest Pauli allowed $SU(3)$ representation 
($(\lambda,\mu)=(0, 4)$) by 66$\%$.
However, $Sp(2,R)_z$ representation can be a better description; 
the squared overlap between the ground state and $\Lambda=5/2$ state, 
whose band head is $(\lambda,\mu)=(0,4)$, is 93$\%$.

\begin{figure}[htbp]
\includegraphics[width=86mm]{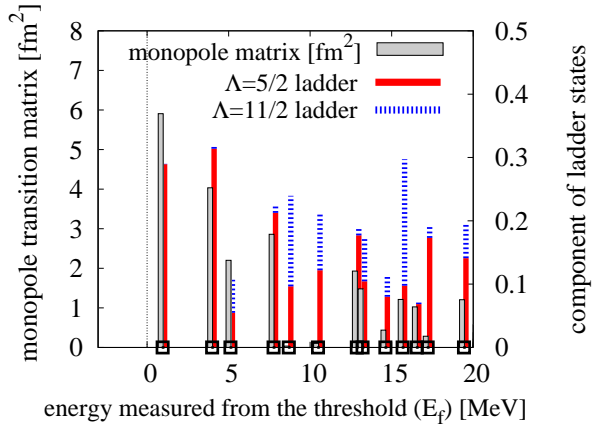}
\caption{(Color online) The relation between the monopole matrix from the ground state (left vertical axis) and components of ladder states of $Sp(2,R)_z$ algebra (right vertical axis) for each $0^+$ state.
The squares show the eigenvalues of the Hamiltonian within the bound state approximation.
}
\label{fig:E0E}
\end{figure}

Next, we discuss the monopole transition matrix element (proton part) 
 from the ground state to excited states 
with the energies of $E_f$ measured from the threshold
 as shown in Fig. \ref{fig:E0E} (left vertical axis). 
The obtained value of $\sim$5.9 fm$^2$ to the second $0^+$ state
 just above the threshold energy (calculated as $E_f = 0.96$ MeV)
 shows good agreement with the experimental value ($5.4\pm0.2$ fm$^2$).
Furthermore, we find correlation between the monopole transition
 strength and a $\Lambda$ component in each excited state
 (right vertical axis of \mbox{Fig. \ref{fig:E0E}}).
Here, $\Lambda=5/2$ (red) and $11/2$ (blue) specify the components of the lowest and the second ladder states for $\mu=4$ in each excited state.
From this figure, we can find that the excited states 
which have large monopole strengths dominantly contain 
components of ladder states with the same $\Lambda$ value as the ground state ($5/2$). 
On the other hand, we can see the tendency that
 the monopole matrix becomes small when the excited states
 dominantly have the components of higher ladder states such as $\Lambda=11/2$.
This is one clue to understand the
 correlation between the $\Lambda$ value of the excited states
 and the monopole transition strength from the ground state.

\begin{figure}[htbp]
\includegraphics[width=86mm]{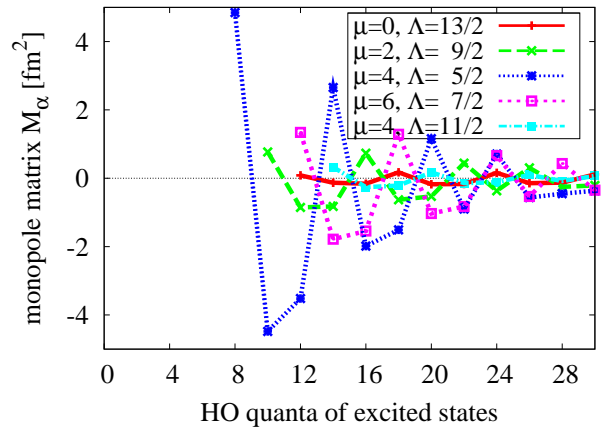}
\caption{
(Color online)  Decomposition of {$M_{\alpha}$}
 in the contribution of each $Sp(2,R)_z$ basis state. 
For a given $\mu$, 
contribution of ladder states with the smallest $\Lambda$ values
 are shown ($(\mu, \Lambda)=$ (0, 13/2) (red), (2, 9/2) (green), (4, 5/2) (blue), (6, 7/2) (purple) and (4, 11/2) (sky blue)). 
}
\label{fig:E0dens}
\end{figure}

In order to understand the above-mentioned behavior of 
the monopole transition with respect to $\Lambda$, 
we expand the monopole matrix $E0(0_1^+\to 0_k^+)$ as
\begin{eqnarray}
E0 (0^+_1 \to 0^+_k) &=& \langle \Phi^{0^+}(k)|\hat{E0}|\Phi^{0^+} (gs) \rangle  \nonumber \\
  &=& \sum_{\alpha}d^{0^+}_{k,\alpha}\langle w^{0^+}_{\alpha}|\hat{E0}|\Phi^{0^+}(gs) \rangle,
\end{eqnarray}
where $\alpha$ again shows an abbreviation of $N(\lambda,\mu)\Lambda$.
At first, we take notice on the matrix element 
 $M_{\alpha}\equiv \langle w^{0^+}_{\alpha}|\hat{E0}|\Phi^{0^+}(gs) \rangle$.
In Fig. \ref{fig:E0dens},
The contribution of each $Sp(2,R)_z$ basis state 
for $M_{\alpha}$ is shown. 
For a given $\mu$, 
the contribution of ladder states with the smallest $\Lambda$ values 
 are shown ($(\mu, \Lambda)=$ (0, 13/2) (red), (2, 9/2) (green), (4, 5/2) (blue), (6, 7/2) (purple) and (4, 11/2) (sky blue)). 
We find that $(\mu, \Lambda)=$ (2, 9/2), (4, 5/2) and (6, 7/2)
 states have large contribution for $M_{\alpha}$. 
The main reason comes from the fact that the monopole operator
 carries only two quanta and components of the ground state
are concentrated in the $\Lambda=5/2$ state. 
The contribution of other $\mu$ and $\Lambda$ states, 
e.g., ($\mu, \Lambda$)=(0, 13/2) (red line) and (4, 11/2) (sky blue) 
are less than 1.0 fm$^2$ at each HO quanta $N$.
\begin{figure}[htbp]
\includegraphics[width=86mm]{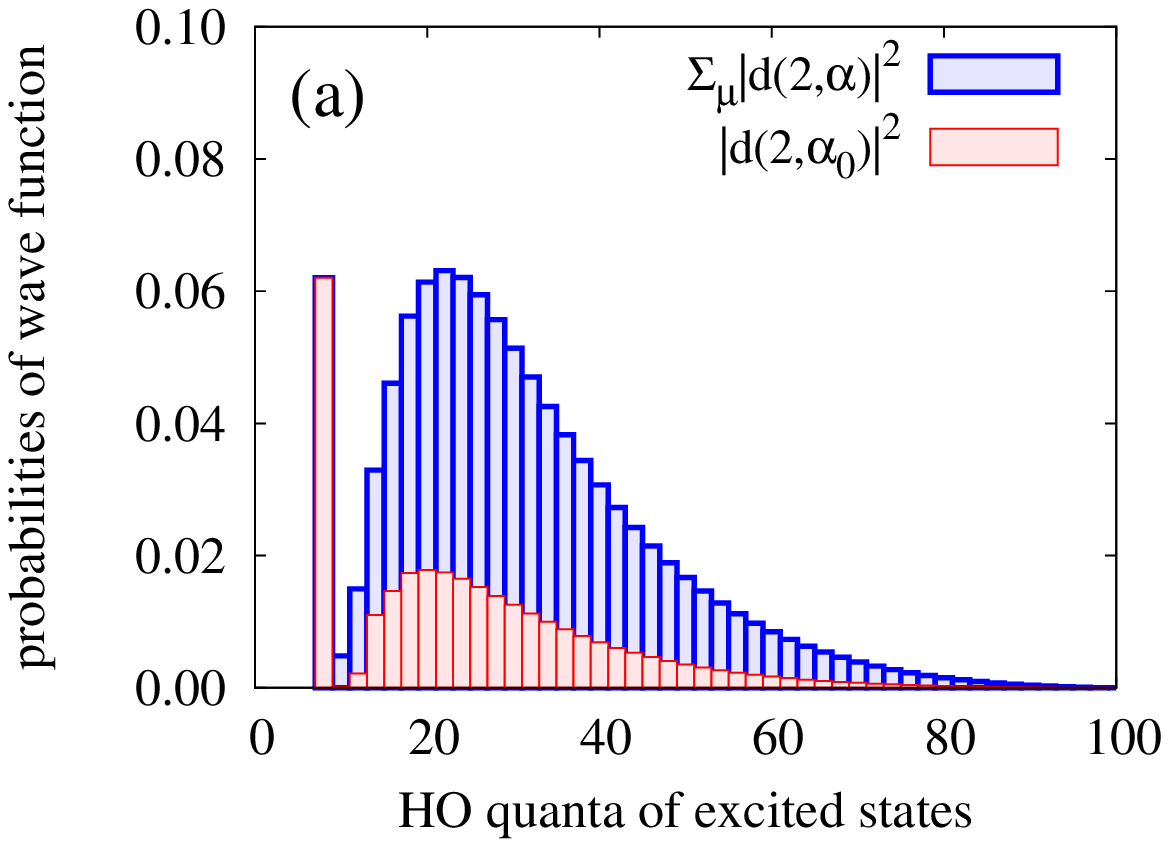}
\includegraphics[width=86mm]{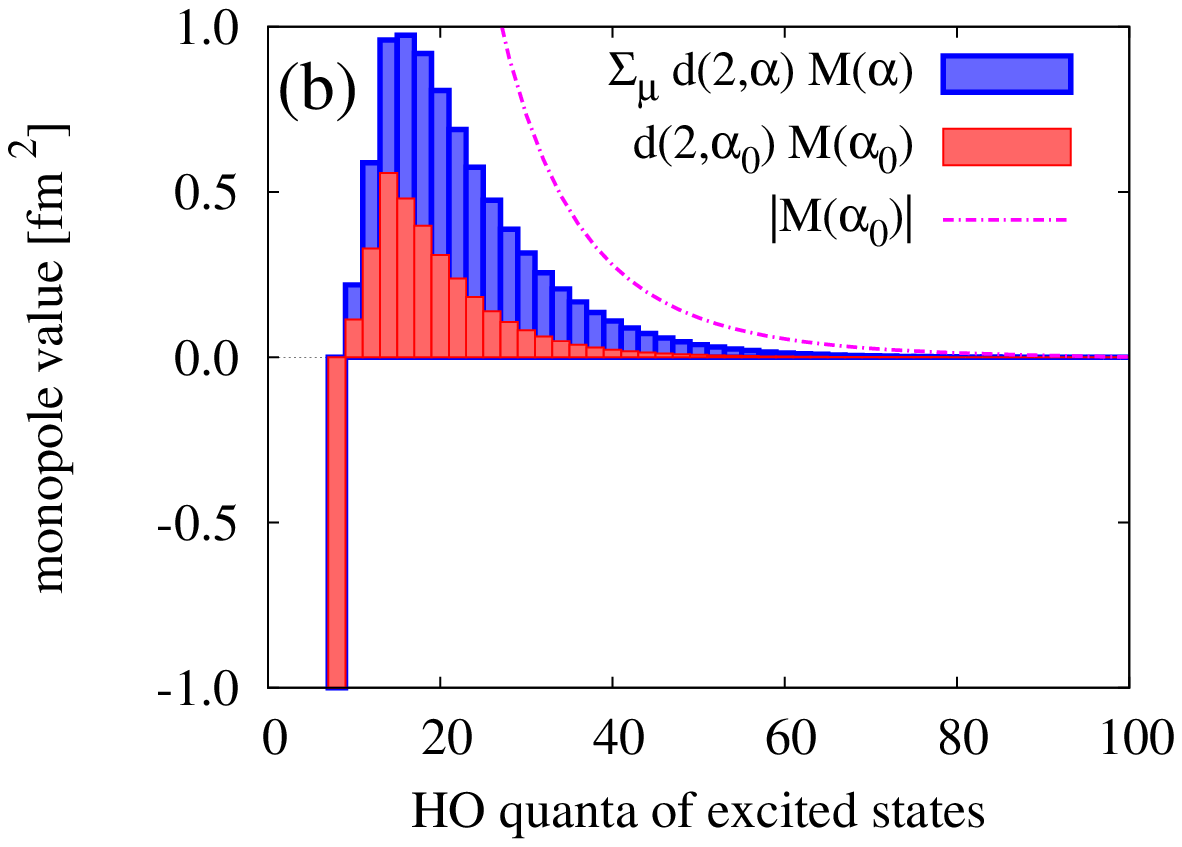}
\caption{
(Color online) The relation between the wave function and monopole matrix for the second $0^+$ state 
($E_f$=0.96 MeV) with respect to HO quanta $N$.
In the upper panel ((a)), the light red box show the component of wave function.
The light blue box shows $\sum_{\mu}|d_{2,\alpha}|^2$, 
and the solid one shows the component of $|d_{2,\alpha_0}|^2$, 
where $\alpha_0$ stands for $(\mu,\Lambda)=(4, 5/2)$.
In the lower panel ((b)), the red and blue box show $M_{\alpha}$ value
 multiplied by the coefficient of the wave function.
The blue box shows $\sum_{\mu}d_{2,\alpha}M_{\alpha}$, 
and red one shows the component of $d_{2,\alpha_0}M_{\alpha_0}$.
The dot dashed purple line shows the $|M_{\alpha_0}|$ value.
Here, $\alpha_0$ shows an abbreviation of $N(\lambda,\mu)\Lambda=N(\lambda,4)5/2$ quanta.
}
\label{fig:transwave-1}
\end{figure}
\begin{figure}[htbp]
\includegraphics[width=86mm]{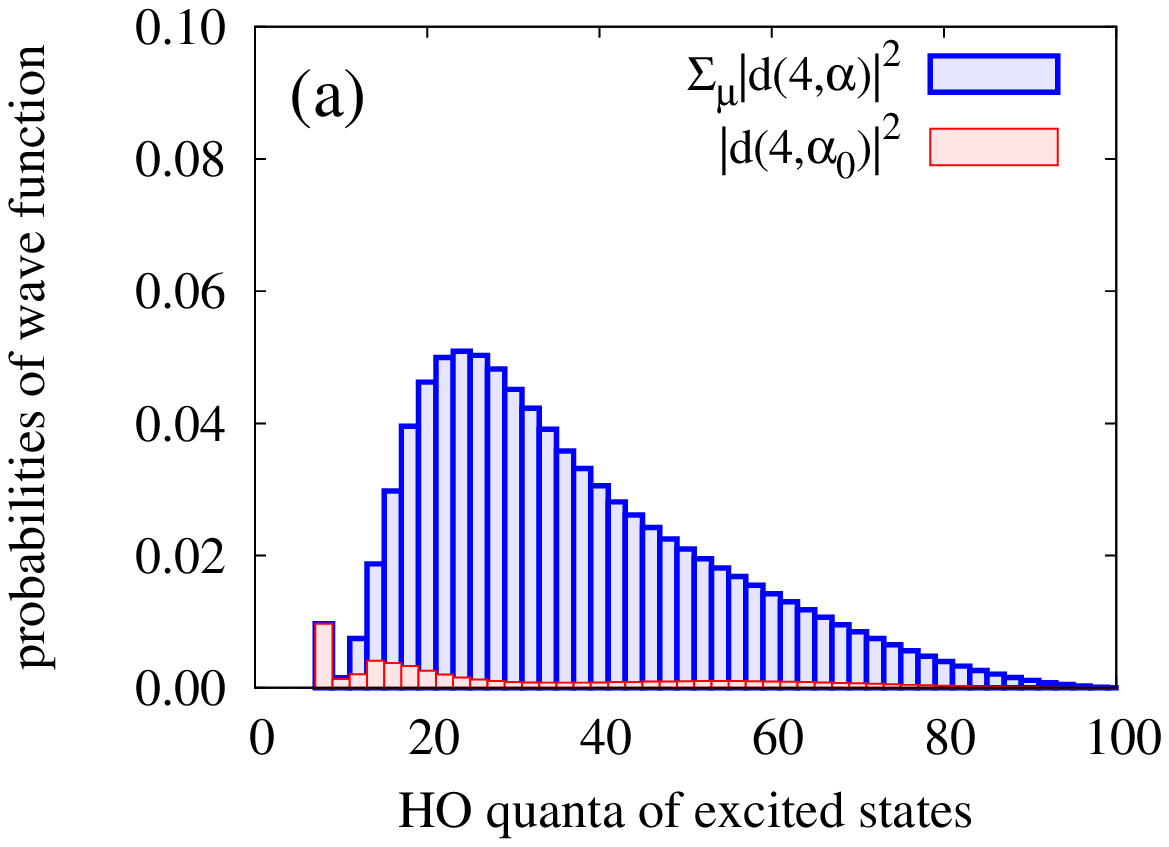}
\includegraphics[width=86mm]{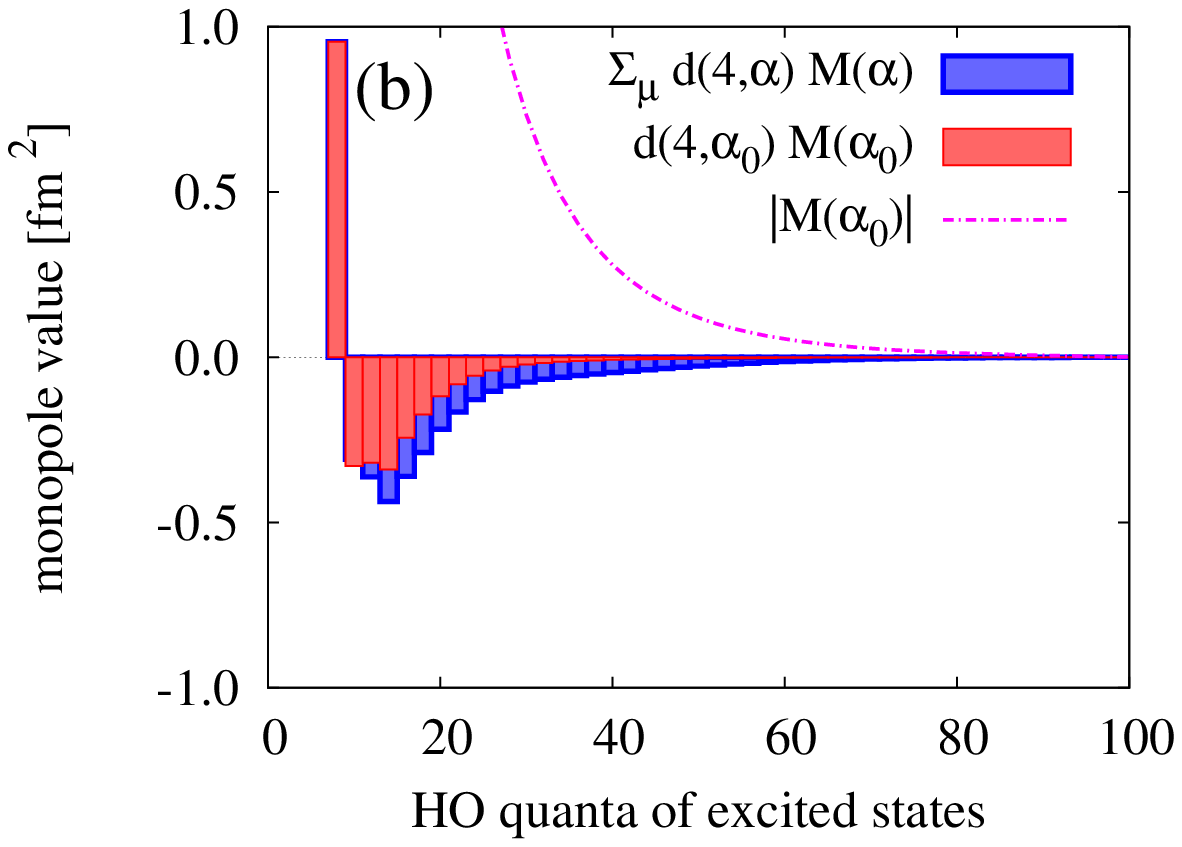}
\caption{
The same analyses as Fig. \ref{fig:transwave-1} for 
the $0^+$ state at $E_f$=5.12 MeV.
}
\label{fig:transwave-2}
\end{figure}
\begin{figure}[htbp]
\includegraphics[width=86mm]{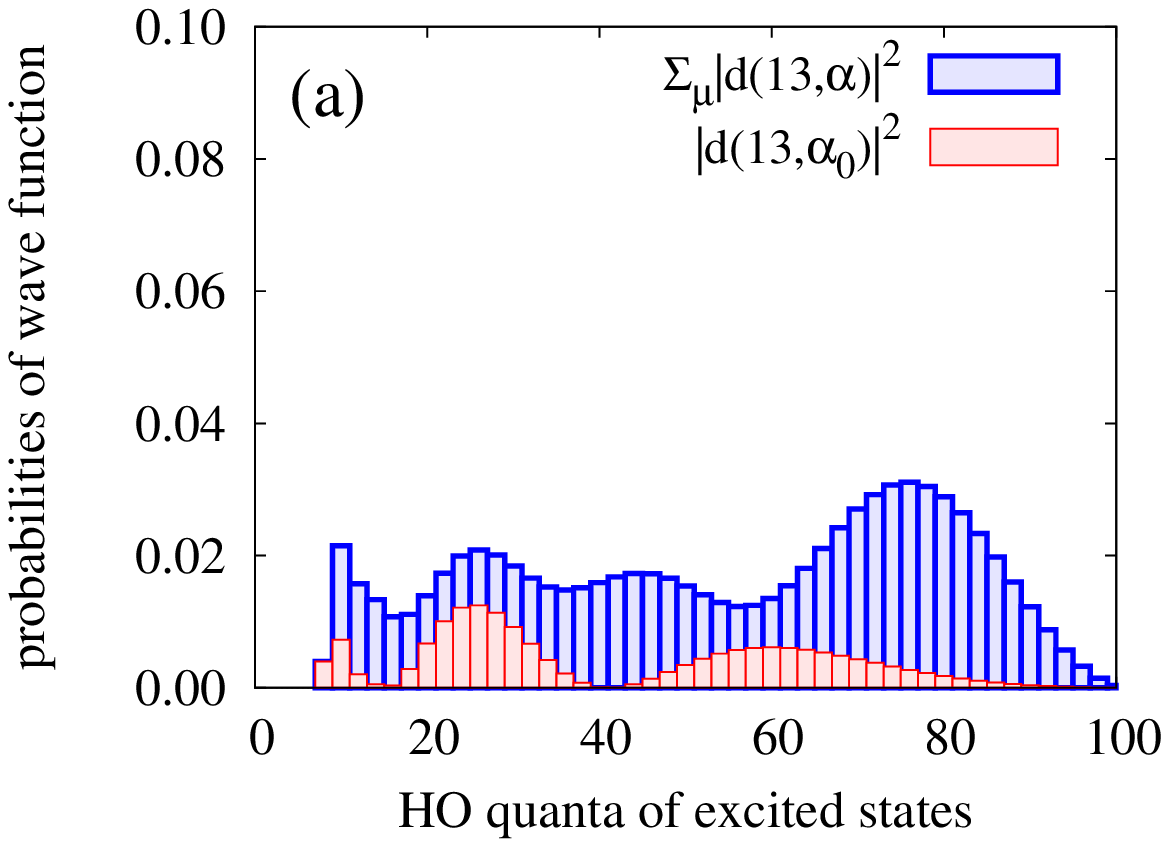}
\includegraphics[width=86mm]{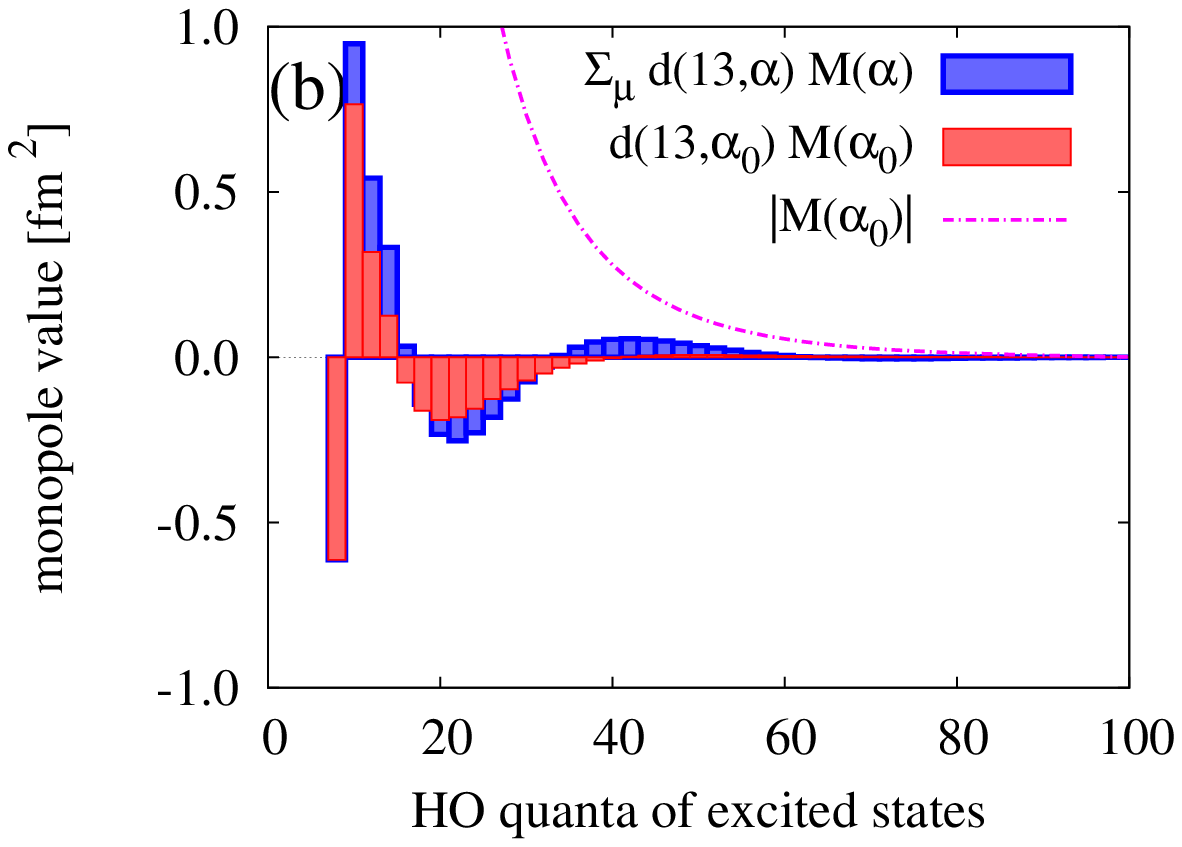}
\caption{
The same analyses as Fig. \ref{fig:transwave-1} for 
the $0^+$ state at $E_f$=17.21 MeV.
}
\label{fig:transwave-3}
\end{figure}
The overall behavior of the monopole transition strength
 is governed by this $M_{\alpha}$ value.
However, 
the detail structure of $E0 (0^+_1 \to 0^+_k)$
 varies depending on the wave function of the excited states.
Therefore, next we discuss the 
relation between the matrix element $M_{\alpha}$
 and the coefficients $d_{k,\alpha}$.

In Fig. \ref{fig:transwave-1}, 
we depict the wave function of the second $0^+$ state (calculated at $E_f=$ 0.96 MeV)
and monopole strength from the ground state.
The following values,  
$\sum_{\mu}|d_{k,\alpha}|^2$ (light blue bars) and 
$|d_{k,\alpha_0}|^2$ (light blue bars) are shown in Fig. \ref{fig:transwave-1} (a), 
while 
$d_{k,\alpha}M_{\alpha}$ (blue bars), 
$d_{k,\alpha_0}M_{\alpha_0}$ (red bars) and
$|M_{\alpha_0}|$ (dot dashed purple line) 
are shown in Fig. \ref{fig:transwave-1} (b).
Here, $\alpha_0$ shows an abbreviation of $N(\lambda,\mu)\Lambda=N(\lambda,4)5/2$ quanta.
Since the small $\Lambda$ states are found to be important (in Fig. \ref{fig:E0dens}),
 here $\Lambda$ (in $\alpha$ and $\alpha_0$) is
 set to be the smallest for a given $\mu$. 

From this figure, we can see which part of the wave function is
 important for the monopole transition strength.
For example, the HO quanta $N$ of the second $0^+$ state ($E_f=$0.96 MeV)
 ranges up to $N\sim60$ (Fig. \ref{fig:transwave-1} (a)).
The important $N$ values can be determined by $d_{k,\alpha}M_{\alpha}$ (red and blue bars) value,
 which shows that the HO quanta up to $N\sim40$ 
coherently contribute to the monopole value (Fig. \ref{fig:transwave-1} (b)).

We can also investigate the transition to even higher excited states.
The transition to the $0^+$ state at $E_f=$ 5.12 MeV
is analysed in Fig. \ref{fig:transwave-2} (a) and (b).
 This state has only small contribution of the $\alpha_0$ state (light red bars).
Even in such case, the small contributions of $d_{k,\alpha}M_{\alpha}$ (light blue bars) create
certain amount of the monopole matrix when they are summed over the HO quanta $N$,
 which is similar to the case of the second $0^+$ state.
The $d_{k,\alpha_0} M_{\alpha_0}$ value (red bars) and $\sum_{\mu}d_{k,\alpha} M_{\alpha}$
 (blue bars) almost overlap with each other,
which suggests the importance of the $\alpha_0$ configuration ($N(\lambda,\mu)\Lambda)=N(\lambda, 4)5/2$)
for the monopole transition strength.

In some of excited states
 ($E_f=$ 3.99, 7.76, 8.69, 10.48, 12.86, 14.61, 15.67, 16.54, 17.21 and 19.43 MeV), 
 the slope of wave function strongly depends on the HO quanta $N$.
As shown in Fig. \ref{fig:transwave-3} (a) and (b),
 the wave function of the $0^+$ state at $E_f$= 17.21 (MeV)
 has clear nodes (light red bars) and they cause cancellation of
the monopole strength (red bars).
 Therefore, the resultant monopole matrix becomes small.
The transition to the states at the energies of 
$E_f=$ 8.69, 10.48, 14.61, 15.67 and 19.43
 MeV from the threshold also shows similar behavior (see Fig. \ref{fig:E0E}).
These states are related to the continuum solution,
 which will be discussed in the next subsection.

We notice that $N$ distributions of wave functions
 are also calculated by FMD (Fermionic molecular dynamics) method \cite{Chernykh}.
The difference between the peak position of the second $0^+$ states of $^{12}$C 
in $N$ (principal quantum number) between the present result and FMD comes from
the definition of $N$.
Our definition is the total principal quantum number, 
while the FMD one is the excitation of principal quantum number from
the lowest shell model state ($N=8$).
If we take into account this shift due to the difference of the 
definition of $N$, both results are quite consistent.
Our peak for the second $0^+$ state around $N=20$ correspond to the peak around $N=16$ in FMD.
The state at $E_f=$ 3.99 MeV has double peaks around $N=16$ and $N=58$.
In the FMD calculation, such double-peak structure appears for the third $0^+$ state
(around $N=$ 14-16 and 52-54).

\subsection{Energy levels and properties of each state}
\begin{figure}[htbp]
\includegraphics[width=8.6cm]{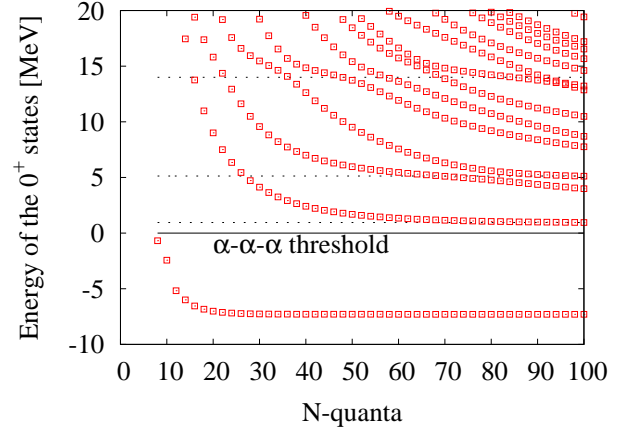}
\caption{(Color online) Energy convergence of the 3$\alpha$ system
 with respect to the increase of the $N$-quanta for the model space ($N_{max}$).
The dotted line (black) shows the stationary points with respect to $N$,
 which are candidates for the resonance states.
}
\label{fig:nconv}
\end{figure}

\begin{figure}[htbp]
\includegraphics[width=8.6cm]{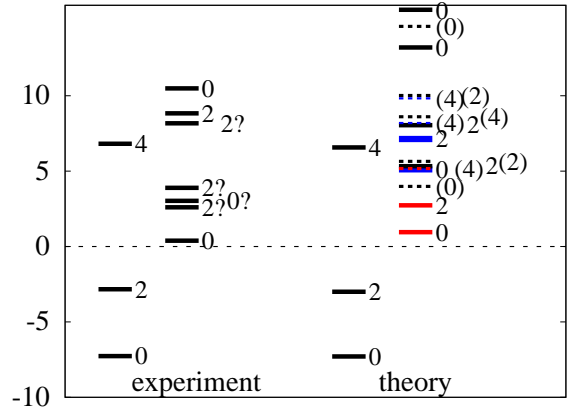}
\caption{
(Color online) Energy spectra of $^{12}$C (positive parity) measured from the threshold}
\label{fig:c12lev}
\end{figure}

In the last subsection, we discussed 
 there is a tendency that states with components of lowest $\Lambda$ are mainly
excited when the monopole operator acts to the ground state.
From this analysis, we can confirm the close relation between 
the symplectic structure and the monopole strength.
However, we must keep in mind
 that not all of states which have large monopole transitions survive
 as resonance states when we impose correct boundary condition.
The extraction of the resonance solution can be performed by drawing energy convergence with
 respect to the increase of the maximum HO quanta of the model space, $N_{max}$.
As shown in Fig. \ref{fig:nconv}, the obtained states show the 
behavior of quasi-stationary solution at the energies of $E_f=$ 0.96 MeV, 5.12 MeV and 14.00 MeV from
the threshold.
These states are candidates for the resonance states.
This is consistent with the previous work in Ref. \cite{Kurokawa}.
The obtained candidates for the resonance states after this treatment are
shown in Fig. \ref{fig:c12lev} together with the bound states.
The left and right spectra correspond to the experimental and theoretical ones, respectively.
The location of the theoretical ground band levels ($0^+$, $2^+$ and $4^+$)
 are fitted to the experimental ones
by adjusting the strength of the three-body interaction given in Eq. (\ref{eqn:3body}).
The excited $0^+$, $2^+$ and $4^+$ states are calculated 
using the same strengths of the three-body interaction as those for the 
ground band states.
We can see a reasonable agreement with the experiment levels the same as 
in the previous calculations \cite{Kurokawa, Yamada-1}.
Here, the dotted lines with the parentheses ($J$) 
show the levels which are obtained as bound state approximation 
but do not show the behavior of stationary solutions by the analysis of Fig. \ref{fig:nconv}.

The property of each level is characterized by red and blue colors. 
Above the threshold, the red colored states,
0$^+$ (0.96 MeV), 2$^+$ (2.73 MeV) and 4$^+$ (5.17 MeV), 
have gas-like nature of three $\alpha$ clusters,
while the blue colored states,
 0$^+$ (5.12 MeV), 2$^+$ (7.12 MeV) and 4$^+$ (8.19 MeV and 9.83 MeV), 
 have considerable amount of linear-chain configurations.
\begin{table}[htbp]
\begin{center}
\caption{Properties of $^{12}$C levels.  The root mean square radius ($R_{r.m.s.}$ (fm)) and 
squared overlap of each state with $(\mu,\Lambda)$ configuration (right column).
The states with the parentheses ($J^\pi$) are obtained as bound state approximation 
but do not show the behavior of stationary solutions.
}
\label{tab:state12C}
\begin{tabular}{cc|c|ccc}
\hline
\hline
  &          &               &          &  $(\mu,\Lambda)$ &      \\
$E$ (MeV) & $J^\pi$  &  $R_{r.m.s.}$ (fm) & (0,13/2) & (2, 9/2) & (4, 5/2)  \\
\hline
-7.29 & 0$^+$ & 2.39 & 0.00  & 0.02 & 0.93   \\
-3.00 & 2$^+$ & 2.45  & 0.00  & 0.03 & 0.91   \\ 
 6.57 & 4$^+$ & 2.82  & 0.02  & 0.05 & 0.80   \\ 
\hline
 0.96 & 0$^+$ & 3.61  & 0.17  & 0.21 & 0.29   \\
 2.73 & 2$^+$ & 3.95  & 0.26  & 0.22 & 0.22   \\ 
 5.17 & (4$^+$) & 4.28  & 0.26  & 0.20 & 0.22   \\
\hline
 5.12 & 0$^+$ & 3.92  & 0.45  & 0.07 & 0.05   \\ 
 7.13 & 2$^+$ & 4.29  & 0.30  & 0.09 & 0.15   \\ 
 8.19 & (4$^+$) &  4.30   & 0.32  & 0.10 & 0.12   \\ 
 9.83 & (4$^+$) &  4.63   & 0.25  & 0.12 & 0.14   \\ 
\hline
\hline
\end{tabular}
\end{center}
\end{table}
These characters are deduced from the calculated root mean square radii
 ($R_{r.m.s.}$) and probabilities of each $(\mu, \Lambda)$ configuration 
listed in Table \ref{tab:state12C}.
The gas-like states are characterized by the large $R_{r.m.s.}$ value, and since
 the wave function is dilutely distributed,  
it has components of various $(\mu, \Lambda)$ configurations.
 For instance, the $2_2^+$ state ($E_f$= 2.73 MeV) is considered to have 
the gas-like nature.
Although a candidate has been reported \cite{Itoh}, 
the excited states of the Hoyle state
have not been experimentally confirmed.

On the other hand, the linear-chain states are characterized by 
large overlap with $\mu=0$ configurations.
The $0^+$ state at $E_f=$ 5.12 MeV obtained within the present
 framework contains the characteristics of linear-chain configuration.
We can see that the amount of the linear-chain component decreases as $J$ increase.
Moreover, the stationary point of energy convergence indicates that 
the linear-chain structures tend to have relatively large decay widths
 than the gas-like states.
Therefore, the clear rotational band structure cannot be seen in the present calculation.

\section{Summary}
In this paper, we have studied the relation between the monopole transition strength 
of $^{12}$C and the special algebraic structure 
to investigate the large strength including the one for $^{12}$C ($0^+_1 \to 0^+_2$).
Here, we have focused on the similarity of the monopole operator 
and the generators of the $Sp(2, R)_z$ algebra.
The model space is constructed based on the $Sp(2, R)_z$ algebra,
 and the ladder states were generated from the band head states
 given by the $SU(3)$ representation.

We have found that the large contribution for the monopole transition strength
 can be explained from the properties of the generators of $Sp(2, R)_z$ and the ground state. 
We have been able to discuss the mechanism that the monopole strengths are
 closely related to the $\Lambda$ value of the final states. 
Here, the importance of the $\Lambda$ ladder state
which is the same as the ground state ($\alpha_0$) 
has been discussed. 
We found that 
the overall behavior of the monopole strength is given
 by the amount of $\alpha_0$ configuration. 
However, the detailed value is sensitive to the properties of the wave function, where we 
have seen these values as a function of the $N$-quanta of harmonic oscillator.
We have also seen that the mechanism appears even in the linear-chain like $0^+$ state 
where the small amount of $\alpha_0$ configuration exists.

We have also checked the stability of these states to 
select the candidates for the resonance states.
For this purpose, we have investigated the behavior of 
the energy convergence with respect to the $N$-quanta of harmonic oscillator. 
We have also analysed whether the obtained states have gas-like or linear-chain structure, 
and the candidate for the excited Hoyle state ($2^+$) has been found.

Since our wave functions are constructed from purely Pauli allowed states,
the applicability for further analyses is quite large. 
For instance, applying non Hermitian formalism by taking correct boundary 
condition based on complex scaling method (CSM) \cite{ABC1, ABC2} is feasible.
In the forthcoming paper, we will construct the formalism which can be combined with CSM. 
The present analysis is an important first step for the analysis along this line.

\end{document}